\documentclass[journal, twocolumn]{IEEEtran}
\usepackage{graphicx} 
\usepackage{xspace}
\usepackage{upgreek}
\usepackage{siunitx}
\usepackage{cite}

\providecommand{\comment}[1]{}
\providecommand{\um}{$\upmu$m\xspace}

\usepackage{fixltx2e}
\usepackage{url}
\begin{document}


\title{\LARGE Optimizations of GaAs Nanowire Solar Cells}
\author{\IEEEauthorblockN{Anna H.\ Trojnar,\IEEEauthorrefmark{1}\IEEEauthorrefmark{2} Christopher ˜E.\ ˜Valdivia,\IEEEauthorrefmark{1} Ray ˜R.\ ˜LaPierre,\IEEEauthorrefmark{3} Karin ˜Hinzer,\IEEEauthorrefmark{1}\IEEEauthorrefmark{2} and Jacob J.\ Krich\IEEEauthorrefmark{2}\IEEEauthorrefmark{1}} \\
\IEEEauthorblockA{\IEEEauthorrefmark{1}SUNLAB, School of Electrical Engineering and Computer Science, University of Ottawa, Ottawa, Ontario, K1N 6N5, Canada}\\
\IEEEauthorblockA{\IEEEauthorrefmark{2}Department of Physics, University of Ottawa, Ottawa, ON, K1N 6N5, Canada}\\
\IEEEauthorblockA{\IEEEauthorrefmark{3}Department of Engineering Physics, McMaster University, Hamilton, Ontario, L8S 4L7, Canada}}

\maketitle

\begin{abstract}
The efficiency of GaAs nanowire solar cells can be significantly improved without any new processing steps or material requirements. We report coupled optoelectronic simulations of a GaAs nanowire (NW) solar cell with vertical p-i-n junction and high band gap AlInP passivating shell. Our frequency-dependent model facilitates calculation of quantum efficiency for the first time in NW solar cells. 
For passivated NWs, we find that short-wavelength photons can be most effectively harnessed by using a thin emitter while long-wavelength photons are best utilized by extending the intrinsic region to the nanowire/substrate interface, and using the substrate as a base. These two easily implemented changes, coupled with the increase of NW height to 3.5~\um with realistic surface recombination in the presence of a passivation shell, result in a NW solar cell with greater than 19\% efficiency.
\end{abstract}


\section{Introduction \label{sec:intro}}
\IEEEPARstart{O}{ver} the past decade there has been rapid improvement in periodic nanowire (NW) arrays for photovoltaic applications. Such NW solar cells (NWSC) are attractive due to their potential for high efficiency, low cost, and low materials utilization. Crucial for solar cell applications, nanowire arrays have excellent light absorption \cite{Tsakalakos07, Zhu09a, Hu07, Krogstrup13, Cao10, Yao14, Cao09, Lin09b,Kupec09} 	
due to a self-concentrating effect despite their small cross-sectional area and low geometric filling factor. An optimized NW array has higher broadband absorption than thin film cells of the same height \cite{Kelzenberg10, Fountaine14a, Lin09b, Huang12a, Madaria12} and achieve performance of equivalent thin-film designs \cite{Wallentin13, Garnett10}. The decreased material usage offered by NW solar cell designs is of particularly high importance as solar cells increase to the terawatt scale.

The small diameter of the NW allows for dislocation-free strain relaxation at the NW surface, which enables use of lattice-mismatched materials to enhance photoconversion efficiency, removing the lattice-matching constraint of planar multi-junction solar cells \cite{Kavanagh10}. 
Similarily, NWs can be grown on inexpensive lattice mismatched substrates such as Si or glass to further reduce cell cost. \cite{Dhaka12, Mohseni13, Feng08,Martensson04,Nguyen11a, Tomioka09,Shin11, Munshi14, Bae14, Russo-Averchi15}.

The main drawback of NWSCs is their high surface-to-volume ratio, making them vulnerable to surface recombination in the absence of proper passivation  \cite{Huang14a,Li15a, LaPierre11a, Mariani13}. A radial p-n junction NWSC design  \cite{Yao14, Kayes05, Hu12, Wang14, Huang14a, Li15a, Yoshimura13, Czaban09, Mariani13, Colombo09, Mariani11} offers the possibility of improved carrier collection, and decoupling of the optical and electrical thicknesses of the active material; however, sidewall contacts bring another set of challenges, by introducing additional Fermi level pinning and diffusion of the contact material into the nanowire during annealing \cite{Kim97}.

Interest in axial-junction NWs has been renewed \cite{Huang14a, Yao14,  Huang12, Aberg16} since the demonstration that passivated sidewalls can have higher performance than sidewall contacts\cite{LaPierre11a}, and the record experimental efficiency for a NWSC is currently 15.3\% \cite{Aberg16}. Axial-junction NWs are also suitable for multi-junction designs. 
Passivation methods for III-V materials are well developed \cite{Hasegawa08, Olson89}, with high bandgap materials shown to well passivate vertical GaAs NWs \cite{Chang12, Huang14a, Chia13}.

Optimizing the design of NWSCs requires accurate modeling of their optical and electrical properties. Much previous work has considered detailed simulations of the wave optics of absorption in NWs but used simple analytical models for the electrical properties, or simple geometric optics with more realistic electrical simulations \cite{Lin09b, Kupec09, Kayes05, Hu12, LaPierre11a, Yu12}. 
These uncoupled models provide insights into correlations between geometry of NW arrays and their absorption or transport properties; however only combined optical and electrical simulations can give quantitative guidance in design parameters for NWSCs \cite{Yao14, Huang14a, Huang12,Li15a, Wang14}. Such combined models are new enough that there are still significant improvements in NWSC design that can be achieved without additional processing steps or material requirements. We show that even the  the best-performing NWSCs \cite{Aberg16} can achieve significant improvements by changing the thicknesses of their p-i-n regions. 

Our coupled optoelectronic device model consists of two steps. First, we perform wave optics calculation of the wavelength- and spatially-resolved optical generation profile in the NW, followed by electrical simulations with a drift-diffusion based device simulator. Depending on the type of the study, we either perform electrical simulations for each wavelength independently, or we use optical generation integrated over the wavelengths to simulate AM1.5D illumination. Previous opto-electronic models use time-domain optical simulations, which do not give frequency-resolved information. Our frequency-domain model additionally allows extraction of wavelength-dependent solar cell properties such as internal and external quantum efficiency, which provide key insights into device performance.

We show that the best performing NWSCs have large intrinsic regions where most of the carrier generation occurs, with substrates acting as a base, and thin emitters. Optimizing the intrinsic length can improve efficiency by over 6\% (absolute) under AM1.5D illumination at 1 sun intensity, and the emitter optimization can improve efficiency by over 3\% (absolute).  
The long intrinsic region improves internal quantum efficiency (IQE) for long wavelength photons while the thin emitter improves IQE for short wavelength photons. A trade off between absorption and surface recombination needs to be taken into account during the NW height optimization. Increasing the NW height leads to increased absorption, but without a passivation shell, the cell efficiency saturates despite increased height. 
These proposed improvements to emitter and intrinsic region doping are easily implemented and should significantly improve device efficiencies. They demonstrate the power of frequency-domain coupled optoelectronic calculations to provide key insights to optimize nanowire device performance.

\section{Models and Methods}
\label{sec:models}

\begin{figure}
\includegraphics[width=0.49\textwidth]{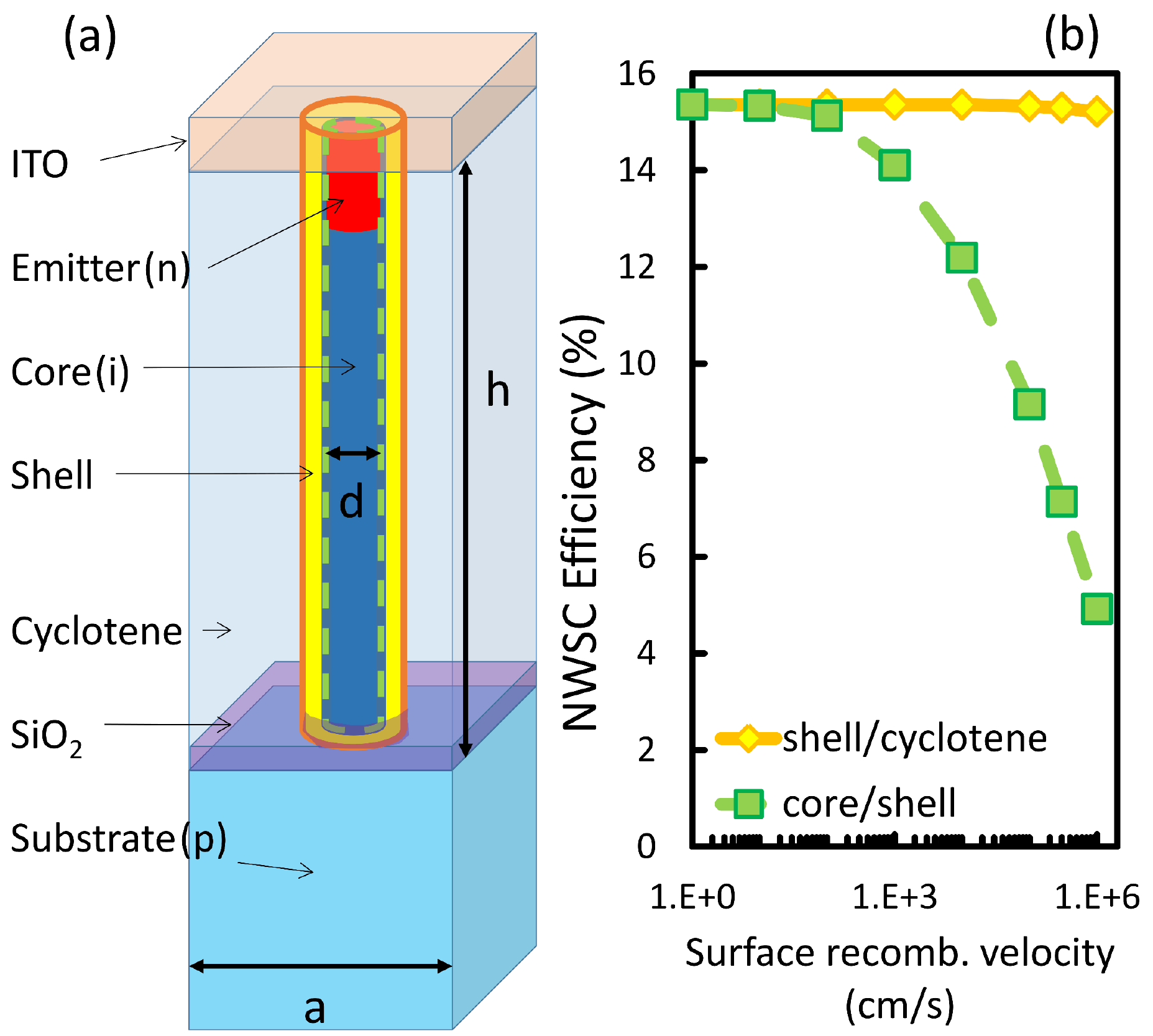} 
\centering
\caption{(a) Side view of a single unit cell containing one NW with the NW core - shell interface in green, and NW shell - cyclotene interface marked orange. (b) NWSC efficiency with 600~nm period, 300~nm diameter, 1.3~\um height, and initial doping (as shown in Table \ref{table_geom}) as a function of surface recombination velocity $S$ at the core - shell surface (green) and shell - cyclotene surface (orange).}
\label{fig_rec}
\end{figure}

We investigate square arrays of vertical GaAs NWs, shown in Fig.~\ref{fig_rec}(a), with square period $a$, core diameter $d$ and cylindrical symmetry. The cell design contains a vertical p-i-n junction with the NW surface passivated by an AlInP shell. At the base of the NW, the core is surrounded by a thin layer of SiO$_2$. The array is planarized with the dielectric polymer cyclotene and contacted at the top by a transparent conducting indium tin oxide (ITO) layer. The NWs have height $h$ and are grown on a GaAs substrate.

\begin{table}
\renewcommand{\arraystretch}{0.9}
\caption{Model parameter ranges for nanowire array geometry and doping.}
\label{table_geom}
\centering
 \begin{tabular}{lccc} 
 \hline\hline
\bfseries Parameter & \bfseries Symbol & \bfseries Value  \\ 
\hline
NW array period [nm]& $a$ & 250-600  \\ 
NW core diameter [nm]& $d$ & 100-360   \\
NW height [nm]& $h$ & 1000-3500   \\
Emitter thickness [nm]& -- & 10-400  \\
Intrinsic thickness [nm]& -- & 50-1260  \\
Substrate thickness [nm]& -- & 1000  \\
Shell thickness [nm]& -- & 20  \\
ITO thickness [nm]& -- & 300  \\ 
Surf. rec. vel. [cm/s] & $S$ & 0--3E5   \\ 
n-Emitter doping [cm$^{-3}$]& --  & 3E18  \\
p-Intrinsic doping [cm$^{-3}$]& -- & 2E16 \\
p-Substrate doping [cm$^{-3}$]& -- &3E18   \\
n-Shell doping [cm$^{-3}$]& -- & 5E15  \\
 \hline\hline
\end{tabular}
\end{table}

Our coupled optoelectronic device model consists of two successive calculations. First, using the frequency-domain optical model, we calculate the optical generation rate within a NW and substrate. Second, the obtained optical generation is used as an input for drift-diffusion based electrical simulations of the transport properties of the NWSC.  


Since the NWs are subwavelength in size, accurate absorption simulation requires consideration of wave optics. 
The absorption properties of the NW array with a square period are obtained by solving the time-harmonic wave equation for the electric field distribution in the 3D simulation domain (see Fig.~\ref{fig_rec}(a)) using the COMSOL Radio Frequency module and the Electromagnetic Waves Frequency Domain interface. 
Frequency domain calculation offers the ability to study wavelength-dependent phenomena and allows us to extract the IQE of the NWSC designs for the first time. 
The optical properties of the materials are defined by specifying wavelength-dependent refractive index $n$ and extinction coefficient $k$ for each material taken from bulk properties \cite{Adachi99,Hu12}. 

In order to simulate broadband absorption of the NWs, the 300-900 nm wavelength range is divided into 120 equally-spaced frequency bins and weighted by the AM1.5D illumination spectrum (900 W/m$^2$). The optical simulation is performed once for each frequency bin, each characterized by the center frequency $f_i$. 

Due to the cylindrical symmetry of the NW, one incident circular polarization is sufficient to simulate unpolarized solar radiation \cite{Yao14}. We inject a $\sigma_+$ polarized plane wave at normal incidence 500 nm above the ITO layer, which allows the model to resolve reflections from the ITO surface. To reduce the computational cost, our model has only a 1 \um thick substrate, which suffices to absorb nearly all photons that produce carriers that can be collected in the device. To eliminate reflections from the bottom surface we place an absorbing port at the bottom boundary of the substrate domain. 
The square periodic NW array and optical coupling between neighboring NWs are simulated by application of periodic boundary conditions for electromagnetic radiation to the vertical boundaries at all sides of the simulation cell. 

The optical calculation determines the spatially and spectrally resolved steady state electric field $E(f_i,\mathbf{r})$ for each frequency $f_i$. 
The optical generation rate at each point {\bf r} is then
\begin{eqnarray}
	G(\mathbf{r})=\sum_i \frac{\varepsilon_0}{\hbar}|E(f_i,\mathbf{r})|^2 n(f_i)k(f_i)	       
\end{eqnarray}
where $\varepsilon_0$ is the permittivity of free space and $\hbar$ is the reduced Planck constant \cite{Wang14, Yao14}. 
After an interpolation between the different meshes of the optical and electronic simulations, described in Appendix \ref{app:OptGen}, optical generation is then passed into the electrical device model.

It has been shown that photon recycling is much weaker in the NWSC than in planar GaAs solar cells \cite{Wang14}, hence we do not take into account this effect in our electrical simulations, which avoids the need for computationally expensive self-consistency in the electrical and optical simulations. 


The transport in NWSC devices was simulated using TCAD Sentaurus, a commercial drift-diffusion based device simulator from Synopsys Inc. 
The main equation solver, which solves the coupled Poisson equation for electrostatic potential and drift-diffusion equations for electron and hole concentration, uses as an input the optical generation calculated in the optical simulation. 
The circular nanowire cross-section allows the electrical calculation to be performed in cylindrical coordinates, neglecting the small asymmetry in optical generation due to the square lattice of NWs (for details see Appendix \ref{app:OptGen}). 
TCAD Sentaurus takes into account the important semiconductor quantities such as Fermi-Dirac statistics for carrier concentrations, electron and hole effective masses \cite{Levinshtein96}, their corresponding effective conduction and valence band densities of states, doping-dependent mobilities \cite{Sotoodeh00}, thermionic emission, generation, and recombination phenomena (radiative \cite{Piprek03}, SRH (Shockley-Read-Hall) with doping dependent lifetimes, and Auger \cite{Levinshtein96}), and surface recombination at material interfaces. 

Reference \citenum{Huang14a} reported that surface recombination on the exterior interface of the shell does not significantly affect the NWSC performance. In agreement, we have also observed that recombination at the shell/cyclotene interface is negligible, since the majority of the photo-carriers are generated in the NW core, and the energetic barriers in the conduction band (CB) and valence band (VB) of the shell prevents their diffusion into the shell. 
We confirm this result, employing surface recombination at the shell/cyclotene interface (see Fig.~\ref{fig_rec}(b) orange-solid line). 
It has been reported, however, that even in NWs well-passivated by either AlGaAs or AlInP shell, surface recombination in the range of $S=1.7$-$5.4 \times 10^3$ cm/s is present at the core/shell interface \cite{Huang12, Lapierre11, Demichel10,Chang12}. In the absence of the passivation, surface recombination velocities near $S=3\times 10^5$ cm/s have been measured \cite{Yao14}.
We find that surface recombination at the core-shell interface (green-dashed line in Fig.~\ref{fig_rec}(a)) can significantly affect the NWSC performance as shown in Fig.~\ref{fig_rec}(b). 

In this work we demonstrate how appropriate choice of nanowire passivation, intrinsic region and emitter thicknesses as well as NW height can significantly improve NWSC efficiency. We emphasize that in the remainder of the manuscript, we consider surface recombination due to imperfect passivation at the core/shell interface, shown with a solid green line in Fig.~\ref{fig_rec}(b).
Throughout this work we present results with four surface recombination velocities: $S=0$, $1\times 10^3$, $3\times 10^3$, $3 \times 10^5$ cm/s, which we call ``ideal,''  ``optimistically passivated,'' ``well-passivated,'' and ``unpassivated''.

\section{Results}
\label{sec:results}
Our main results show that a thick intrinsic region in the p-i-n structure aids collection of carriers produced by long-wavelength light and a thin emitter aids collection of carriers produced by short wavelengths. Together, these simple design modifications significantly improve the predicted device efficiency.

The most important source of the NWSC performance enhancement (6\% abs.) comes from the increase of the intrinsic layer thickness in the p-i-n structure from 50 nm to the whole remaining extent of the NW height, using the substrate as a base. This modification improves carrier collection within the NW due to existence of an electric field through the whole NW core rather than only in a thin layer, as well as enabling collection of carriers generated by long wavelength radiation in the substrate. 

There is also a performance boost due to a decrease of the emitter thickness, caused by a shift of the carrier generation into the intrinsic region. We find that a cell with emitter thickness of 40 nm and well (optimistically) passivated surface has 18.0 (18.8)\% efficiency. Since the same design with 300 nm thick emitter has an efficiency of 14.7 (15.4)\%, the adjustment of only the emitter thickness improves the cell performance of the otherwise optimized design by 3.3 (3.4)\% absolute.

These two optimization changes, coupled with an increase of NW height to 3.5 \um, results in a NWSC with 19.1 (20.3)\% efficiency.

\subsection{Array geometry and doping}
\label{ssec:geometry_dop}
The optical properties of the NWSC depend mainly on three NW array geometry parameters: period, nanowire diameter and height \cite{Yao14,Huang12,Huang14a, Hu12}. Their optimization is crucial and they have been widely studied. 
We investigated the influence of the NW array geometry -- period $a$ and NW diameter $d$ -- on the NWSC performance.  Our results agree with previous work showing that the performance of the solar cells improves with an increase of the array pitch, with $d/a$ constant, until the maximum design performance is reached near $a$=600 nm and $d/a$=0.5. For details see Appendix \ref{app:geometry}.
%

We performed a separate study (not shown) to optimize doping levels, giving the optimal values shown in Table \ref{table_geom}. We consider nanowires of height 1300~nm to minimize growth time while achieving most of the efficiency gains of taller nanowires. Effects of nanowire height are considered in Sec.\ \ref{ssec:height}, which shows that further height increases can give less than 1\% improvement (absolute) in efficiency with realistic passivation. In all below simulations we use this optimized doping, $a$=600~nm, $d$=300~nm, and height of $h$=1300~nm, unless other values are specified.

\subsection{Thick intrinsic regions are best}
\begin{figure}[ht]
\includegraphics[width=0.49\textwidth]{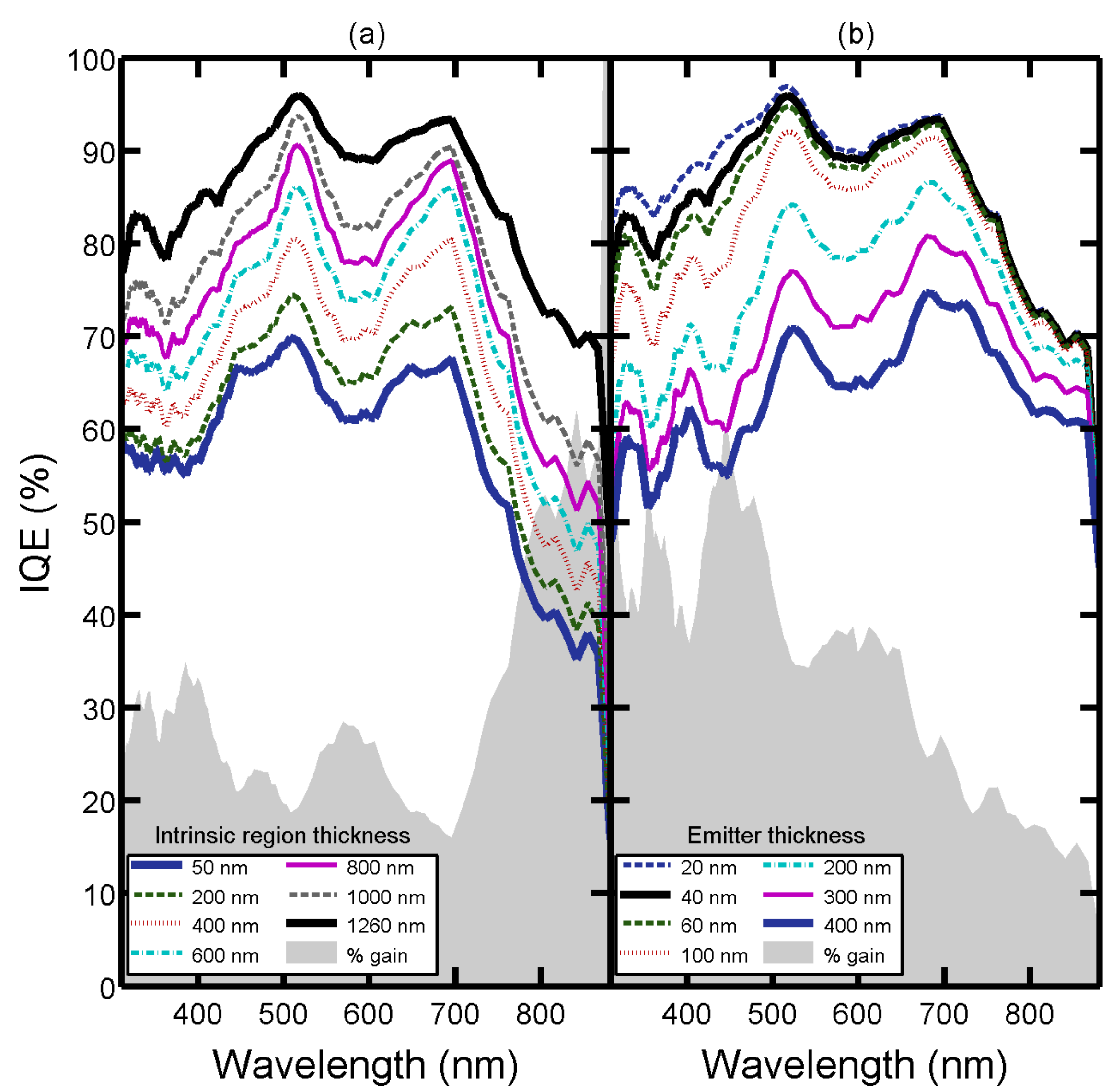}
\centering
\caption{Internal quantum efficiency of NWSC for various (a) intrinsic layer thicknesses and (b) emitter thicknesses, for standard NW geometry with $S=3 \times 10^3$~cm/s. Gray shaded regions show relative IQE improvement when the intrinsic layer thickness increases from 50~nm to 1260~nm or emitter thickness decreases from 400 to 40~nm, respectively. The solid black lines are the same structure in both plots. Thicker intrinsic regions improve collection of carriers generated by long wavelength radiation, deep in the NW. Thinner emitters improve collection of carriers generated by short wavelength radiation at the top of the NW.}
\label{fig_iqe}
\end{figure}

Though many nanowire solar cell devices have been made with p-n structures \cite{Aberg16} or p-i-n structures with thin intrinsic regions \cite{Wang14, Yao14}, we show here the benefits achieved from increasing the intrinsic region thickness. Thick intrinsic regions have been modeled before, but they were not the focus of study \cite{Huang14a}. 
The IQE is the ratio of the number of carriers collected by the solar cell to the number of photons of a given energy absorbed by that cell, and it  provides a good measure of the carrier collection efficiency, decoupled from the photon absorption efficiency. Figure \ref{fig_iqe}(a) shows the IQE of the NWSC as a function of the intrinsic region thickness, which has been varied from 50~nm, as in Ref.\ \citenum{Yao14}, to 1260~nm -- covering the whole extent of the NW core that is not part of the emitter. In these studies, the emitter thickness is 40~nm. The thick intrinsic region is particularly helpful for the longer wavelengths of light, which are absorbed deep in the NW and substrate. 

\begin{figure}[ht]
\includegraphics[width=0.49\textwidth]{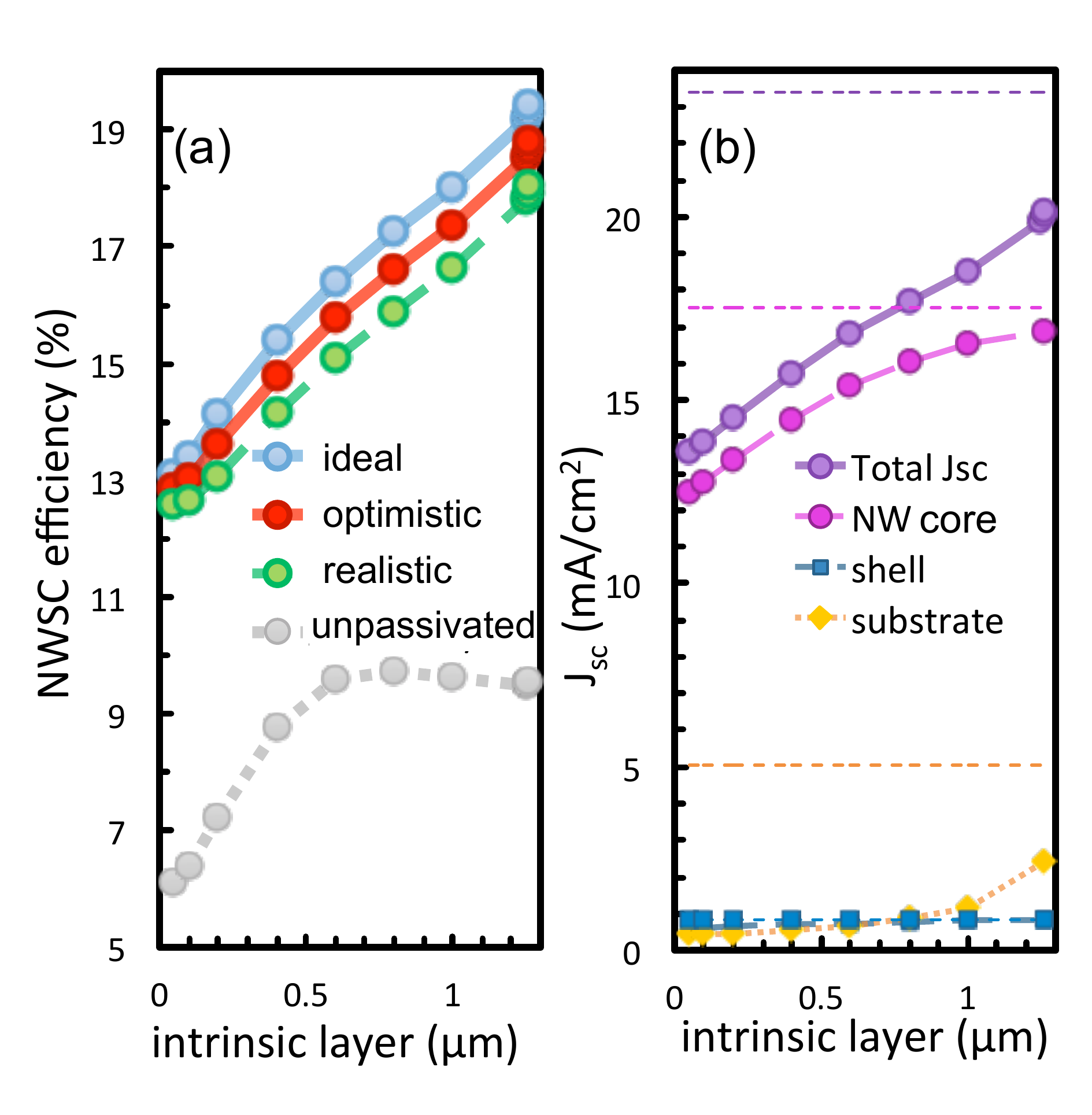} 
\centering
\caption{(a) NWSC efficiency as a function of intrinsic layer thickness for  different $S$. NW height=1300~nm and emitter thickness=40~nm, so maximum intrinsic layer thickness is 1260~nm.
(b) Short circuit current $J_{sc}$ generated in NW core, substrate, shell and the total $J_{sc}$ as a function of the intrinsic layer thickness for realistic S. Dashed horizontal lines show $J_{gen}$, the $J_{sc}$ if all carriers were collected, with the same color code.}
\label{fig_intr}
\end{figure}

NWSC efficiency improves with an increase of the thickness of the intrinsic layer, as shown in Fig.~\ref{fig_intr}(a). This increase is close to linear for all passivated systems, while for unpassivated NWs the efficiency saturates at approximately 9.5\% when the intrinsic layer thickness exceeds approximately $600$~nm. In the unpassivated NW, the carriers generated deep within the core, which could benefit from the extended intrinsic region, are more likely to recombine at the surface than to be collected. Hence, in this case an increase of the intrinsic layer thickness beyond $600$~nm does not improve the carrier collection any further.
We consider the contribution of each region of the NW to the short-circuit current $J_{sc}$ as shown in Figure~\ref{fig_intr}(b). $J_{sc}$ contributions from all regions show improvement as the intrinsic region thickness increases, which confirms improvement in carrier collection.

%
%
Particularly strong improvement in $J_{sc}$ takes place for carriers generated in the NW core as they are extracted more efficiently due to longer minority carrier lifetimes, greater carrier mobilities, and the built-in electric field assisting with photo-generated carrier separation.
The collection of the carriers generated within the substrate by long wavelengths also increases significantly as the intrinsic region thickness reaches the full NW core length, which is clearly visible as an increase of the substrate $J_{sc}$ in Fig.~\ref{fig_intr}(b). 

Figure~\ref{fig_intr}(b) shows the ideal photocurrents $J_{gen}$ (i.e., $J_{sc}$ in the absence of recombination) calculated for each region separately (thin dashed lines).
When the intrinsic region thickness is maximal, $J_{sc}$ of the NW core closely approaches the $J_{gen}$. Both $J_{sc}$ and $J_{gen}$ of the shell are always small due to the high band gap $E_g$ of AlInP.
On the other hand, the long intrinsic region allows collection of carriers produced in the substrate; substrate $J_{sc}$ improves by a factor of 6. 
Despite this improvement, it is clear that the difficulties in carrier collection from the substrate are responsible for the discrepancy between total $J_{sc}$ and $J_{gen}$;  a taller NW can improve carrier collection by shifting absorption to the core from the substrate.

\subsection{Thin emitters are best}
A decrease of the emitter thickness provides a significant source of NWSC efficiency improvement, shifting the carrier generation into the intrinsic region.
In previous work we have shown that the NWSC performance can be significantly improved by reduction of the emitter thickness \cite{Trojnar15}. Here we correct a geometric error in those calculations, as described in Appendix \ref{app:OptGen}, and show that the conclusion is unchanged. 

\begin{figure}
\includegraphics[width=0.225\textwidth]{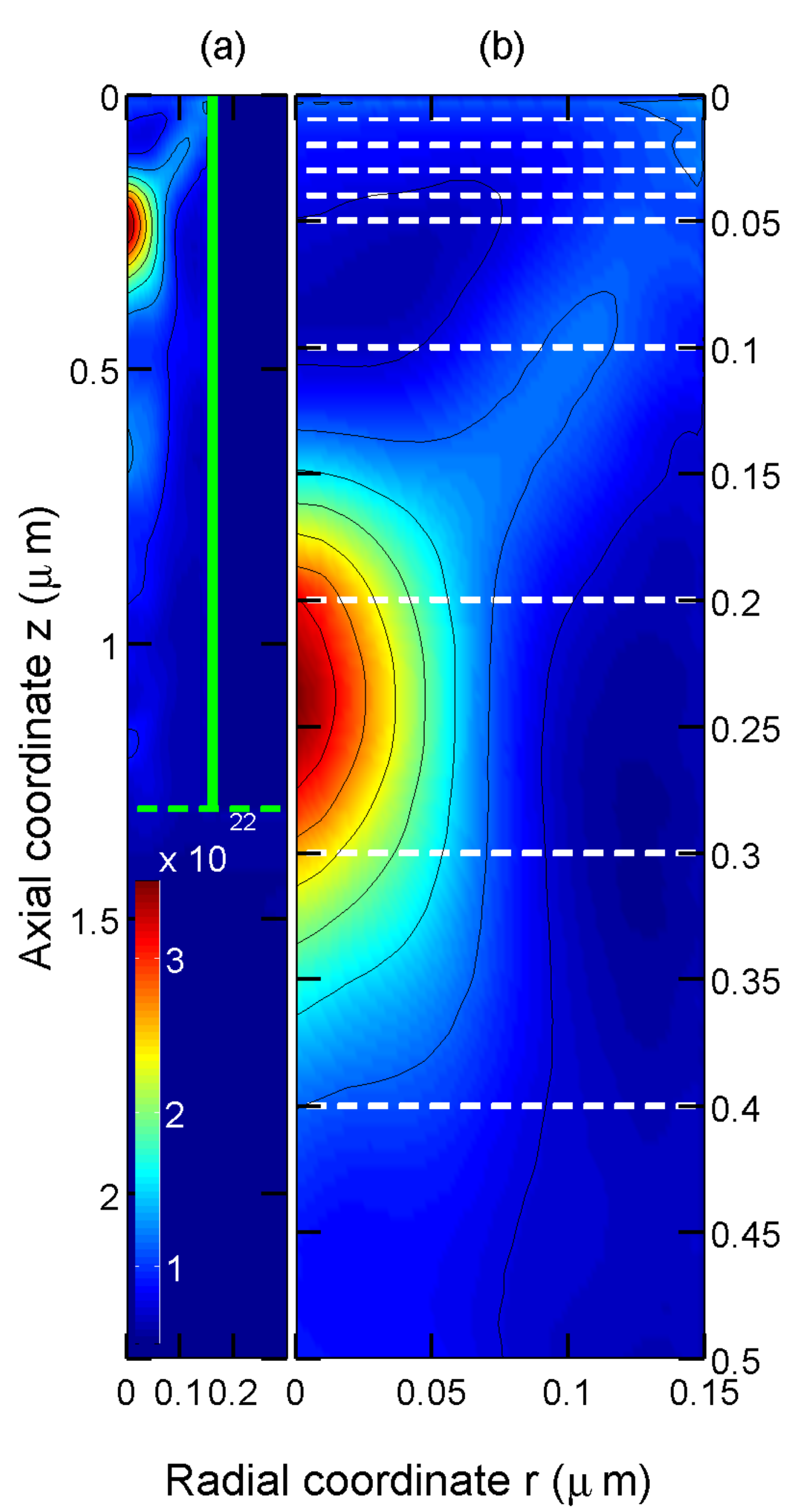}
\includegraphics[width=0.24\textwidth]{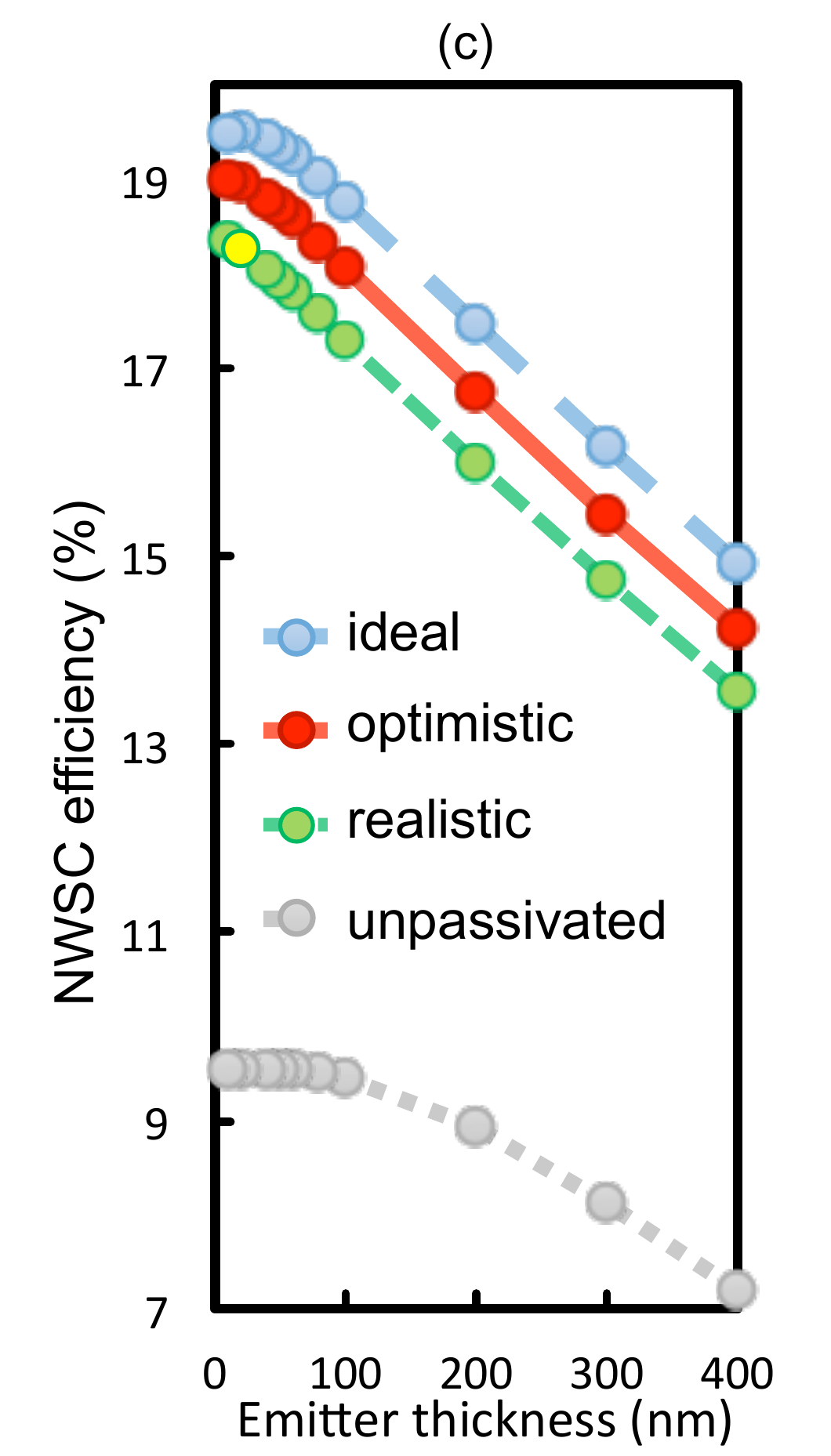}
\centering
\caption{(a) Optical generation in cm$^{-3}$s$^{-1}$ in the NW and surrounding cyclotene of a NW array and substrate of standard geometry. The shell is indicated by a solid-green vertical line. The horizontal dashed line separates the NW from the substrate. (b) Enlargement of (a); dashed horizontal lines indicate simulated emitter thicknesses. (c) Efficiency of the NWSC as a function of emitter thickness, for NW with the standard geometry as in (a), optimized doping and in the presence of four different $S$ as indicated. In all cases, efficiency improves with thinner emitter.}
\label{fig_gen}
\end{figure}

We varied the emitter thickness from 10 to 400~nm for a NW with the standard geometry, specified in Sec.\ \ref{ssec:geometry_dop}. Figure ~\ref{fig_gen}(a,b) shows an optical generation profile in the NW, shell, substrate, and surrounding cyclotene. For the considered NW designs, the position of the peak optical generation is mostly insensitive to a wide range of NW dimensions, appearing 150-220~nm below the top surface.

The NWSC performance increases as the emitter thickness decreases, since most of the carrier generation is moved into the intrinsic region, where carrier collection benefits from longer lifetimes, greater carrier mobilities, and the built-in electric field assists with photo-generated carrier separation as shown in Fig.~\ref{fig_gen}(c). 
Our NWSC standard design with 40~nm thick emitter and with realistic, well-passivated surface $S = 3\times 10^3$ cm/s has 18\% efficiency (distinguished by the yellow marker in Fig.~\ref{fig_gen}(c)), while the ideal efficiency of this design is 19.4\% ($S$=0 cm/s). In all other calculations, we use a design with a 40~nm thick emitter, which is a manufacturable layer thickness. 

When the absorption occurs in the emitter, some carriers recombine at the top contact rather than diffusing to the junction. Figure \ref{fig_iqe}(b) shows that the IQE increases with a decrease of the emitter thickness, with the greatest improvement at short wavelengths, which are absorbed closer to the top of the NW, and hence benefit more from generating carriers in the intrinsic region. 

\subsection{Effects of nanowire height \label{ssec:height}}
\begin{figure}
\includegraphics[width=0.49\textwidth]{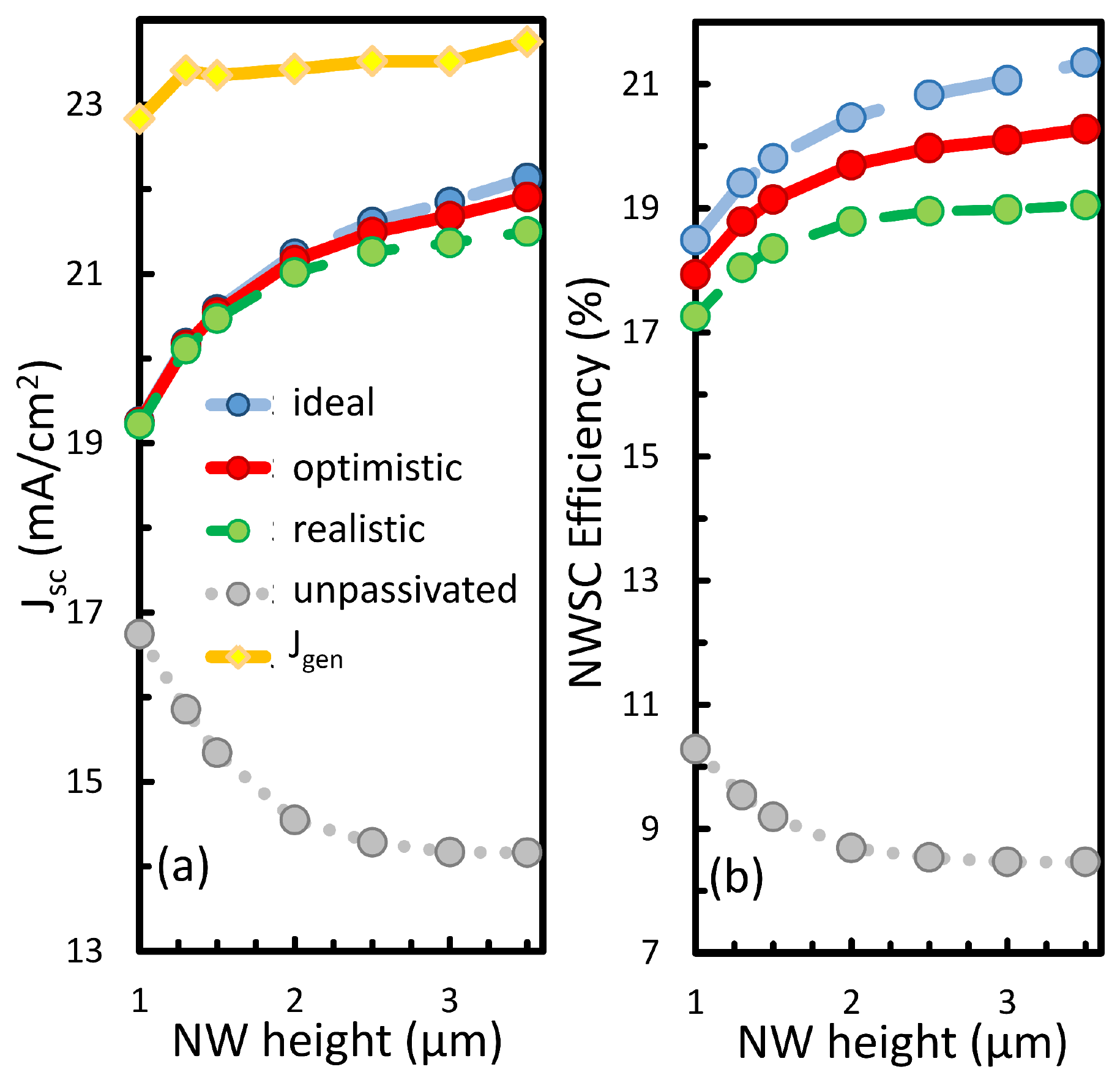} 
\centering
\caption{(a) Short-circuit current $J_{sc}$, and (b) NWSC efficiency with emitter thickness = 40~nm and intrinsic region extending through the NW core, as a function of NW height. Maximum current $J_{gen}$ of the design is shown in yellow. All currents shown here as well as cell performance refer to the whole design, namely NW + substrate. }
\label{fig_h}
\end{figure}

For ideal or well-passivated NWs, an increase of the NW height leads to an increase in the short-circuit current $J_{sc}$ and efficiency (Fig.~\ref{fig_h}(a) and (b), respectively), with the efficiency of our well (optimistically) passivated design reaching 19.1\% (20.3\%). 

For well-passivated cells, the improvement in efficiency comes from increased collection efficiency not from increased $J_{gen}$, since as the NW height increases there are only marginal changes in overall absorption, shown by $J_{gen}$. For tall NWs, more carrier absorption occurs in the NW core where the carrier collection is excellent, instead of the substrate. 
However, one might worry that a tall NW would lose too much current to surface recombination. Figure \ref{fig_h}(a) shows that even for NWs with 3500~nm height, $J_{sc}$ remains close to $J_{gen}$ for well-passivated NWs. 

In the case of an unpassivated NW, both $J_{sc}$ and efficiency decrease with NW height. The decrease of the NWSC performance with the NW height in the presence of high recombination is due to the fact that as the height of NW increases, both the distance that the generated carriers need to travel in order to be collected and the surface area with the high recombination velocity increase. These effects increase the probability of carrier recombination at that surface. This result implies that if proper passivation is impossible, it may be beneficial to decrease NW height in order to achieve higher cell performance. 

\subsection{Loss analysis}
An analysis of losses indicates that the efficiencies described in this manuscript are lower bounds for reasonably achievable NWSC devices. Table \ref{table_loss} shows the losses due to bulk and surface recombination in a NWSC. Both well- and ideally-passivated systems have core/shell recombination lower than total bulk recombination. Appendix \ref{app:electrical_model} contains details of our bulk recombination model, which is standard for bulk GaAs. In the unpassivated case, the loss at the core/shell interface is dominant. For all other cases, the majority of the loss is at the back contact, indicating that carriers generated in the substrate are recombining before finding the nanowire. This loss is an artifact of having only 1~\um of substrate in our model; this substrate thickness is sufficient to model the optical absorption (see Fig.~\ref{fig_gen}(a), which is an expensive 3D calculation; we estimate that increasing the substrate thickness to realistic values would decrease losses in the substrate by approximately 60\%, increasing $J_{sc}$ by 1.5~mA/cm$^2$.  Similarly, a front surface field could further reduce the losses at the emitter/ITO interface. These results imply that realistic devices have the potential to exceed the efficiencies that we have shown in this manuscript.

\begin{table}
\renewcommand{\arraystretch}{1}
\caption{Decrease in $J_{sc}$ in mA/cm$^2$ due to various recombination mechanisms for a NWSC with height $1300$~nm, $40$~nm emitter, and $1260$~nm intrinsic region.}
\label{table_loss}
\centering
 \begin{tabular}{lS[table-format=3.2]S[table-format=3.2]S[table-format=3.2]S[table-format=3.2]} 
 \hline\hline
 &\multicolumn{4}{c}{\bfseries Core/Shell $S$ (cm/s)}\\
 \bfseries Process& \multicolumn{1}{c}{\bfseries 0}  & \multicolumn{1}{c}{\bfseries 1E3} &\multicolumn{1}{c}{ \bfseries 3E3} &\multicolumn{1}{c}{ \bfseries 3E5}  \\ 
\hline
\bfseries Interfaces\\
Core/Shell & 0 & 0.03 & 0.07 &4.38 \\
Emitter/ITO  & 0.54 & 0.54 &0.54&0.53 \\\vspace{2pt}
Back Contact &2.57 & 2.57 &2.57&2.53  \\ 
\bfseries Bulk\\
SRH  & 0.05 & 0.05 &0.05&0.05 \\ 
Auger & 0.007 & 0.007 &0.007 &0.007 \\
Radiative & 0.05 & 0.05 &0.05&0.05 \\
 \hline\hline
\end{tabular}
\end{table}

\section{Conclusion}
In this work we report on optimization of key device parameters such as array geometry, NW doping, emitter thickness, intrinsic layer thickness as well as NW height via coupled optoelectronic simulation. We  find that the most important source of the NWSC performance enhancement (6\% abs.) is the extension of the intrinsic layer thickness from 50 nm to the whole extent of the NW core. This modification improves carrier collection within the NW by extending the region in which an electric field exists and enables collection of carriers generated by long wavelength radiation in the substrate, which acts as a base in this design. Additionally, decreasing emitter thickness improves carrier collection in a short-wavelength range by shifting carrier generation into the intrinsic region, significantly increasing device efficiency.  
These two simple optimization changes coupled with the increase of NW height to 3500~nm with realistic surface recombination results in a NWSC with 19.1\% efficiency. 

\appendices
\section{Optical Generation}
\label{app:OptGen}
\begin{figure}
\includegraphics[width=0.49\textwidth]{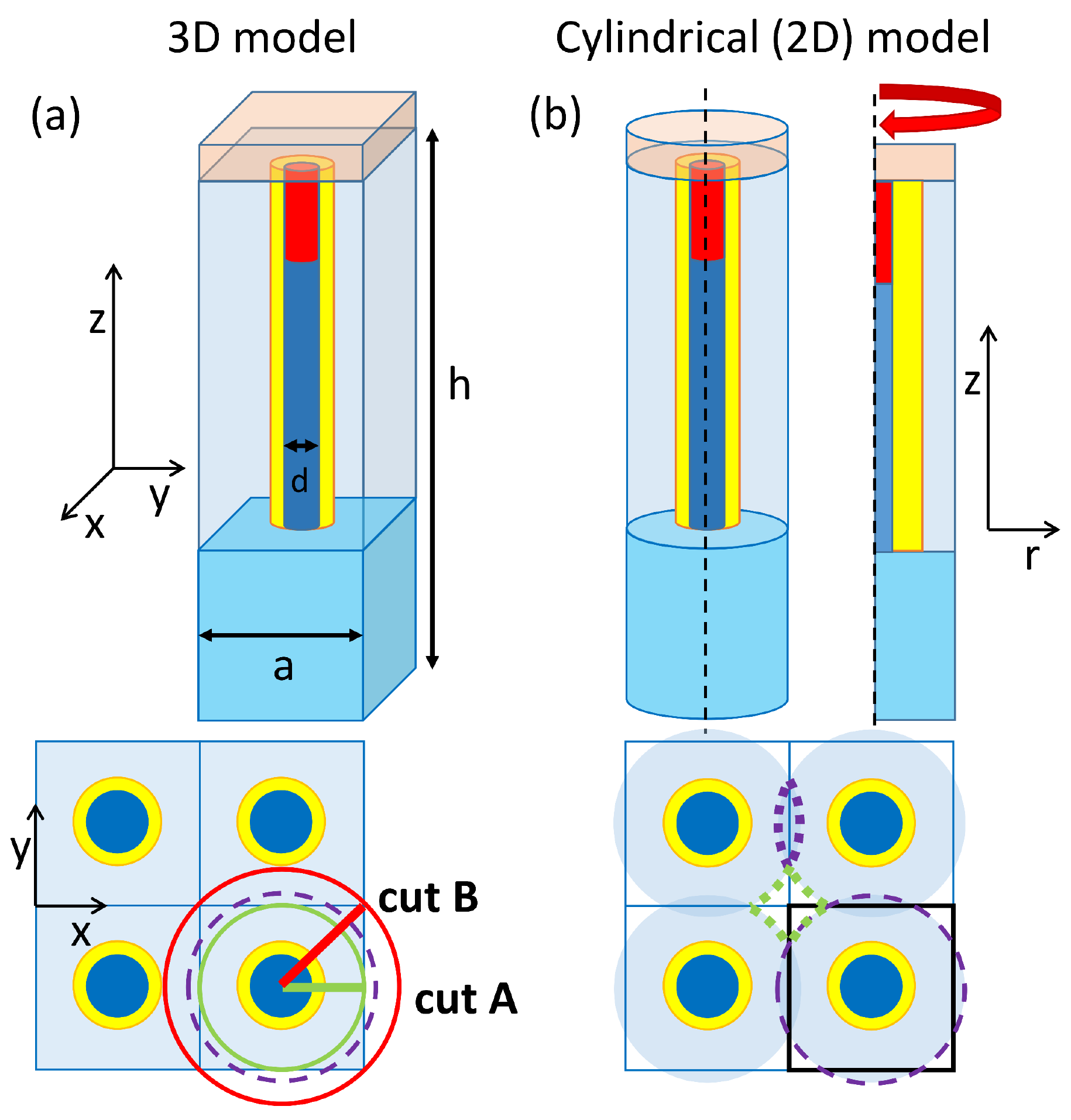} 
\centering
\caption{(a) The 3D simulation structure containing one NW, and a top view of a part of the NW array. Green and red solid lines represent the planes on which $E(f_i,\mathbf{r})$ is extracted, while the circle of corresponding color - showing the equivalent area. Purple-dashed line denotes simulation domain of the electrical model. (b) 3D geometry in cylindrical coordinates and the A simulation plane. A top view of the NW array shows the areas of the simulation domain overlap (purple-dottted lines) and areas that are not simulated in electrical model (green-dotted lines)} 
\label{fig_model}
\end{figure}

We describe here details of the optical model. Due to the square array geometry, the problem does not have a rotational symmetry even for cylindrical nanowires. Deviations from rotational symmetry in the NW optical absorption are not significant and are unlikely to be practically important as carriers can quickly diffuse across the radius of the nanowire; the effect is slightly more significant in the substrate absorption. We describe here the procedure for extracting 2D effective optical generation from the 3D optical simulations. 
Due to the square symmetry of the array we only need the electric field norm $|E(f_i,\mathbf{r})|$ obtained from the 3D simulation within two planes: the {\bf y,z} plane called cut A, that passes though the symmetry axis {\bf z} and a middle of the side wall of the simulation domain, indicated in the top view of Fig.~\ref{fig_model}~(a), and the diagonal {\bf x=y,z} plane (cut B), that passes though the symmetry axis of NW and a corner of simulation domain. We obtain the average electric field distribution needed for further calculations  in a plane through the NW symmetry axis {\bf z}, by averaging the $|E(f_i,\mathbf{r})|$ extracted in A and B cuts.

The electric field norm of all simulation domains contained within the green circle marked in Fig.~\ref{fig_model}(a) and obtained using this method, is a good approximation of the 3D field average, as both A and B planes contain information about those domains. For the domain between the green and red circles the array periodicity is used to extend the plane A data to larger {\bf r}.
It is important to note that this procedure does not affect the absorption properties of the NW, and only impacts the treatment of the substrate absorption. 

In the cylindrical model, all quantities are defined on the $r$,$z$ plane. The radius $R$ must be chosen such that the area of a unit cell is correctly modelled, i.e. $a^2=\pi R^2$.
Our previously reported simulation, used $R=a/2$ which overestimated NWSC efficiencies \cite{Trojnar15}. 
The comparison between these circular areas, and the square unit cell base in the 3D optical model is shown in Fig.~\ref{fig_model}(b).
The approximation of the square by a circle results in the existence of the areas in which $E(f_i,\mathbf{r})$ is either double counted or neglected. These regions contain only substrate and are not expected to strongly affect results.
In the worst case this cartesian to cylindrical transformation procedure produces integrated optical generation input to TCAD Sentaurus that is within 1.5\% agreement with the integrated optical generation extracted from full 3D COMSOL data. Generally the agreement is within 1\%, which translates to 0.3 mA/cm$^2$ uncertainty for the maximum achievable current $J_{gen}$ from the structure. 
\section{Details of electrical model}
\label{app:electrical_model}
Shockley-Read-Hall recombination lifetimes are calculated using the Scharfetter relation
\begin{equation}
\tau_{dop} (N_{A,0},N_{D,0} )=\tau_{min} + \frac{\tau_{max}-\tau_{min}}{1+\left(\frac{N_{A,0}+N_{D,0}}{N_{ref}}\right)^\gamma },
\end{equation}
where maximum lifetime for electrons and holes are set to be $\tau_{max}^e=1\times 10^{-6}$ s, $\tau_{max}^h=2\times 10^{-8}$ s, respectively, and $\tau_{min}^e$=$\tau_{min}^h$=0. 
Together with default Sentaurus parameters for GaAs: $N_{ref}^e$=$10^{16}$ cm$^{-3}$, $N_{ref}^h$=2$\times 10^{18}$ cm$^{-3}$ and $\gamma$=1, these parameters result in SRH lifetimes for electron (hole) of ~10 (20) ns in the emitter, 500 (20) ns in the intrinsic region, and 2 (1) ns in the substrate. 

At the core/shell interface we define surface recombination as
\begin{equation}
R_{surf,net}^{SRH}=\frac{np-n_{i}^2}{(n+n_{i})/s_p +(p+p_{i})/s_n },
\end{equation}
where ${n_{i}}^2 =n_0 p_0$ is an intrinsic carrier concentration, and $n_0$ ($p_0$) are the electron (hole) equilibrium concentrations. 
For convenience, we set a single surface recombination velocity $s_p=s_n=S$ for both electron and holes.

In the electrical simulation we model doping dependent bandgap narrowing, but not in the optical ones, where materials are defined by single $n$, $k$ value set. The effective shrinkage of $E_g$ is caused by existence of band tails within $E_g$, and results in the lowering of the open circuit voltage $V_{oc}$ of the solar cell. On the other hand, there is no need to consider below-bandgap absorption in the optical simulation, since the tail states are localized, and carriers excited to these states do not directly contribute towards the electrical current. We believe this pessimistic combination of assumptions best reflects the real device properties.

\section{Array geometry}
\label{app:geometry}

Three different diameter/pitch ratios $d/a$ = 0.4, 0.5, and 0.6 were considered \cite{Yao14,Huang12,Huang14a} to investigate the influence of the NW array geometry on the NWSC performance.

We varied the surface recombination velocity $S$ with initial doping: emitter $n=8\times 10^{18}$ cm$^{-3}$, intrinsic region $p=1\times 10^{15}$ cm$^{-3}$, base $p=1\times 10^{18}$ cm$^{-3}$, and shell $n=5 \times 10^{15}$ cm$^{-3}$. 

\begin{figure}
\includegraphics[width=0.49\textwidth]{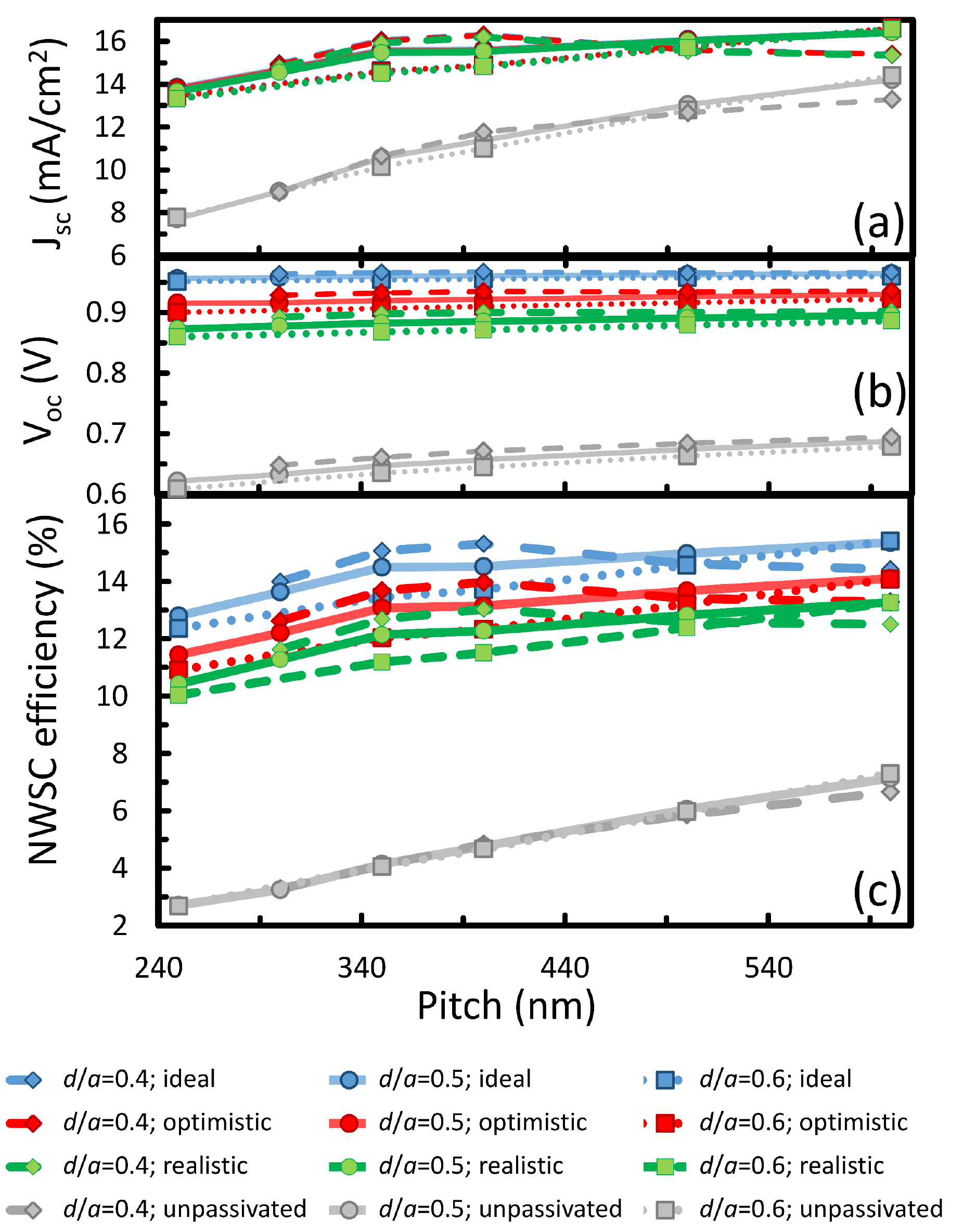} 
\centering
\caption{(a) Short-circuit current $J_{sc}$ of the NW array with substrate, (b) open-circuit voltage $V_{oc}$, and (c) NWSC efficiency as a function of NW array period for three different diameter/period ratios $d/a$=0.4 (dashed-diamond line), 0.5 (solid-circle line), and 0.6 (dotted-square line), The calculations were performed for four representative surface recombination velocities $S$. }
\label{fig_geom}
\end{figure}

Figure~\ref{fig_geom} shows the calculated NWSC (NW + substrate) $J$-$V$ parameters as a function of NW array pitch. The short circuit current $J_{sc}$ shown in Fig.~\ref{fig_geom}(a) generally increases with the NW array pitch. The $J_{sc}$ of well-passivated NWs is close to that of an ideal cell with $S$=0 cm/s, while large surface recombination substantially reduces $J_{sc}$ (grey lines) -- particularly at the small NWSC pitch, resulting in over 43\% $J_{sc}$ decrease. This reduction is not as strong (13\%) for NW arrays with a large pitch. At the same time, open circuit voltage (Fig.~\ref{fig_geom}(b)) of the NWSC is weakly affected by the array geometry, while exhibiting more sensitivity to $S$; even the small surface recombination at well-passivated NWs reduces $V_{oc}$. 

The performance of solar cells with $d/a$=0.5 (solid lines) and 0.6 (dotted lines) improves with an increase of the array pitch. This improvement is greater for NWSC with higher surface recombination velocity, reaching 4.6\% absolute for unpassivated NWs.
We have shown previously \cite{Trojnar15} that for arrays of thinner NWs with $d/a$=0.4, the decreasing volume ratio between photovoltaically active GaAs and surrounding inactive cyclotene with increasing pitch, causes a decrease of ideal photo-induced current $J_{gen}$. This decrease causes a drop in $J_{sc}$ and efficiency  (dashed lines) in Fig.~\ref{fig_geom}(a) and (c). The efficiency of the NW with $d/a$ =0.4 and high $S$ does not follow the trend in $J_{gen}$, since the surface recombination is a dominating loss in carrier extraction. In this case the benefit of increased diameter exceeds the drawback of lower absorption. 

\section*{Acknowledgement}

The authors thank the Natural Sciences and Engineering Research Council of Canada (NSERC) and the Canada Research Chairs program for support, as well as CMC Microsystems for the provision of products and services that facilitated this research, including COMSOL Multiphysics and TCAD Sentaurus. The authors thank  Khalifa M. Azizur-Rahman, and Matthew Wilkins for insightful discussions.

\comment{Aberg and Wallentin both had \% removed from bibliography -- not compiling with them present. Added by hand below.}



\end{document}